# Tunable cavity-enhanced photon pairs source in Hermite-Gaussian mode

Zhi-Yuan Zhou,[1,2,#1] Yan Li,[1,2,#1] Dong-Sheng Ding,[1,2] Wei Zhang,[1,2] Shuai Shi,[1,2]

Bao-Sen Shi,[1,2*] and Guang-Can Guo[1,2]

[1]*Key Laboratory of Quantum Information, University of Science and Technology of China, Hefei,*

*Anhui 230026, China*

[2]*Synergetic Innovation Center of Quantum Information & Quantum Physics, University of Science*

*and Technology of China, Hefei, Anhui 230026, China*

[*]*Corresponding author: drshi@ustc.edu.cn*

The spatial modes of light have grasped great research interests because of its great potentials in optical communications, optical manipulation and trapping, optical metrology and quantum information processing. Here we report on generating of photon pairs in Hermite-Gaussian (HG) mode in a type-I optical parametric oscillator (OPO) operated far below threshold. The bandwidths of the photon pairs are 11.4 MHz and 20.8MHz for two different HG modes respectively, which is capable to be stored in cold Rubidium atomic ensembles. From correlation measurements, non-classical properties of HG modes in different directions are verified by tuning the cavity. Our study provides an effective way to generate photon pairs with narrow bandwidth in high order spatial modes for high dimensional quantum communication.

In modern optical science, the spatial degree of freedom of photon has grasped great research interests for its wide applications in many fields including optical communications [1-3], optical manipulation and trapping [4], optical metrology [5-7] and quantum information processing [8-12], etc. Quantum information processing benefits greatly from improved manipulation of different photonics degrees of freedoms [13, 14]. Various studies of photon's spatial degree of freedoms are motivated that such modes can increase greatly the information carried by photons [15-18].

In quantum communications, quantum memories for orbital angular momentum (OAM) qubit or qutrit states [19-21] and entangled states[22] have been realized. Recently, Laurat's group has demonstrated quantum memory for vector polarized beams [23]. In addition to OAM modes and vector polarized modes, Hermite-Gaussian (HG) mode is another very important spatial mode of the photon, which is eigenmode of an optical cavity. To build quantum memories for HG modes, one need to prepare photonics states in this mode which has a matched bandwidth to couple with the memory. One of the common methods to reduce the bandwidth of photon pairs emitted from spontaneous parametric down conversion (SPDC) in a nonlinear crystal is to use an optical parametric oscillator (OPO) operated far below the cavity's threshold [23-32]. A type-II OPO to generate hyperentanglement continuous-variable in HG mode was reported in ref. [33] recently. So far there is no any report on generating photons pairs in the discrete-variable regime in HG

---

[#1] **These two authors have contributed equally to this article.**

modes using an OPO.

In this letter, we report on generating photon pairs in HG mode in a type-I OPO operating far below threshold. From correlation measurements, we show that the photon pairs has non-classical correlations and the bandwidth of the photon pairs generated is 11.4MHz and 20.8MHz for two different directions of the HG mode. We also show that directions of HG modes can be adjusted by tuning the cavity. Our work provides an effective way to generate photon pairs with narrow bandwidth in high order spatial modes for high dimensional quantum communication.

The Hamiltonian for a type-I OPO operated far below threshold is expressed by [33, 34]:

$$H_{int} = i\hbar\chi \hat{a}_p \hat{a}^\dagger_{s,l} \hat{a}^\dagger_{i,-l} + H.c \quad (1)$$

Here $\hat{a}_p$ ($\hat{a}^\dagger_s$, $\hat{a}^\dagger_i$) is (are) the annihilation (creation) operator for the pump field (signal, idler photon), respectively. $l$ represents the OAM index of the emitted photons, we focus on $l=1$ in this work. For conservation of OAM in the SPDC process, the signal and idler photons have opposite OAM. Therefore, there are two possible ways by which the paired photons are created. The two possibilities are (i) a signal photon is emitted in $LG_0^1$ and an idler photon is emitted in $LG_0^{-1}$, and (ii) a signal photon is emitted in $LG_0^{-1}$ and an idler photon is emitted in $LG_0^1$. Because of the overlap of $LG_0^1$ and $LG_0^{-1}$ in space and time, the $HG_{01}$ and $HG_{10}$ mode are created in the output of the OPO.

The experimental setup is depicted in Fig. 1. In our experiment, the 795nm laser beam from Ti: sapphire laser is divided into two parts by using a polarized beam splitter (PBS). One part of the beam is used for second harmonic generation (SHG) to generate 397.5nm UV beam as the input for the OPO; the other part of the beam is modulated with an electoral optical modulator (EOM) with a frequency of 10.8 MHz, which generates side band for locking the cavity using Pound-Drever-Hall (PDH) method [35]. The Gaussian beam of the input is converted to $LG_0^1$ mode using a vortex phase plate (VPP) before it couples to the OPO.

The OPO cavity consists of two mirrors with a radius of curvature of 80 mm, the input coupling mirror CM2 (high reflection coated at 795nm and high transmission coated at 397.5nm) is attached to piezoactuator for scanning and locking the cavity, and mirror CM1 (4.5% transmission at 795nm and high transmission at 397.5nm) is used as output coupling mirror. A type-I periodically poled KTP crystal(PPKTP, from Raicol crystals, 1mm×2mm×10mm) is placed at the center of the cavity, whose temperature is controlled with a homemade temperature controller.

The OPO cavity and the UV pump beam are mode-matched using a single mode fiber (SMF). The locking beam and the pump beam are combined using a dichromatic mirror (DM). We use a chopper to time-divided between the locking beam and the created photon pairs for detection. The reflected beam from the cavity is rotated by a Farady rotator (FR) and detected with a fast photodiode (PD) for locking the cavity. The generated photon pairs are separated with a half wave plate (HWP) and a PBS, and detected using avalanche photon detector (APD) for coincidence measurement (Timeharp 260, from Pico quanta).

The non-classical correlation of the signal and idler photons is characterized using cross-intensity correlation between them. The normalized cross-intensity correlation is defined as following [28, 36]:

$$g_{jk}^{(2)}(\tau) = \frac{\langle E_j^\dagger(t) E_k^\dagger(t+\tau) E_k(t+\tau) E_j(t) \rangle}{\langle E_j^\dagger(t) E_j(t) \rangle \langle E_k^\dagger(t+\tau) E_k(t+\tau) \rangle} \quad (2)$$

Where indices $j, k \in \{s, i\}$ represent for the signal or idler photon, respectively. The measurements of $g_{jk}^{(2)}(\tau)$ consists of cross-correlation measurement detections between mode $j$ and $k$ at a time delay $\tau$ and the auto-correlation measurement in mode $j$ and in mode $k$.

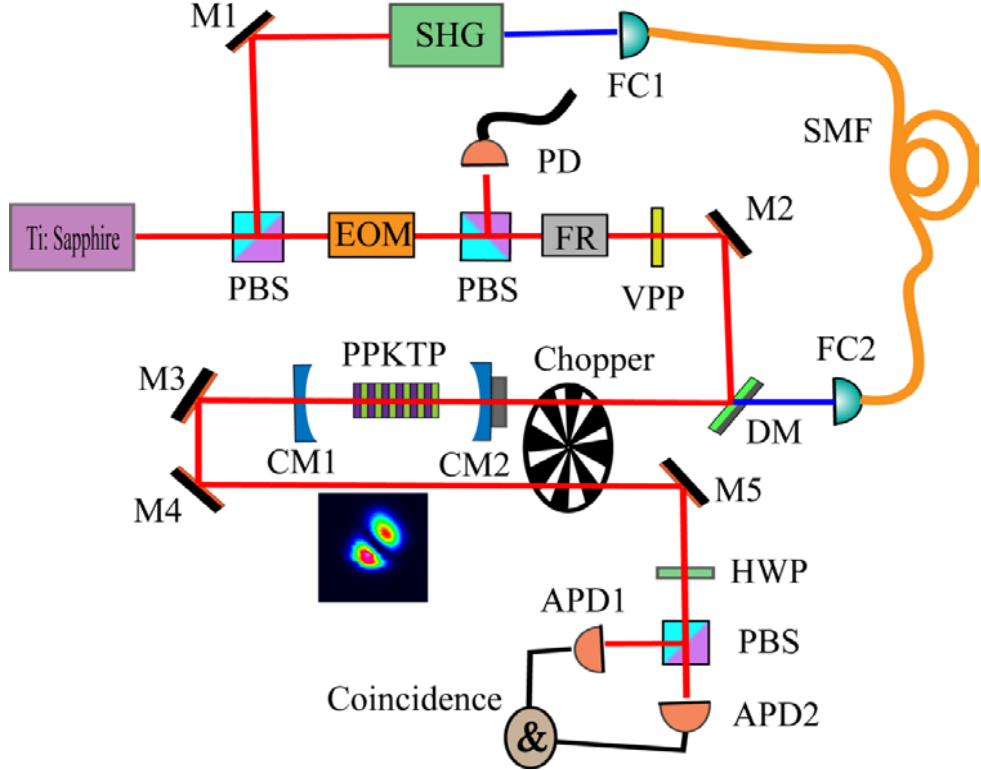

Figure 1. Experimental setup for the experiment. PBS: polarized beam splitter; M1-M5: mirrors; VPP: vortex phase plate; DM: dichromatic mirror; HWP: half wave plate; FC1, FC2: fiber coupler; CM1, CM2: cavity mirrors; SHG: second harmonic generation; FR: Farady rotator; PD: photodiode; SMF: single mode fiber; APD1, APD2: avalanche photon detector.

For measurements of cross-intensity correlation of signal and idler photon in HG mode, we use a pinhole to filter out one petal of the spatial mode and coupling it to single mode fibers. The pump power of the 397.5nm light is 60 µW. The experimental results for HG mode in different directions are showed in figure 2. Figure 2(a) is the results when the cavity mode is tuned to diagonal direction, the spatial shape leaked from the cavity is depicted in the inserted image. The value of the normalized cross-intensity correlation is $g_{s,i}^{(2)}(0) = 7.2$ in a time bin of 0.8 ns, the time for the coincidence measurements is 600 s. The full wave half maximum (FWHM) of the cross correlation time is 19.4ns, and the estimated bandwidth of the photon pairs is 11.4MHz. Figure 2(b) is the result for HG mode in the horizontal directions. The corresponding value of the normalized cross-intensity correlation is $g^{(2)}_{s,i}(0) = 5.7$. The FWHM of the cross-correlation time is 10.6 ns, and the bandwidth of the photon pair is 20.8MHz. $g^{(2)}_{s,i}(0) > 2$ means that the photon pairs have non-classical correlation. The different FWHMs between these two situations are arising from the different mode losses inside the cavity. Curves of the cross-intensity correlation should have comb-like shapes, but because of limited response speed of the APDs and a relative small round trip time (0.94 ns) of photon inside the cavity, the comb-like shape can not be observed. The detailed reasons can be found in ref. [32]. The free spectral range of the cavity is 1.06GHz, the number of the longitude modes which can simultaneously resonance with the cavity for the two cases are 1500 and 2100. The estimated spectral brightnesses of the photon pairs in account for the all losses are 16 (s.MHz.mW)$^{-1}$ and 4.4 (s.MHz.mW)$^{-1}$ respectively.

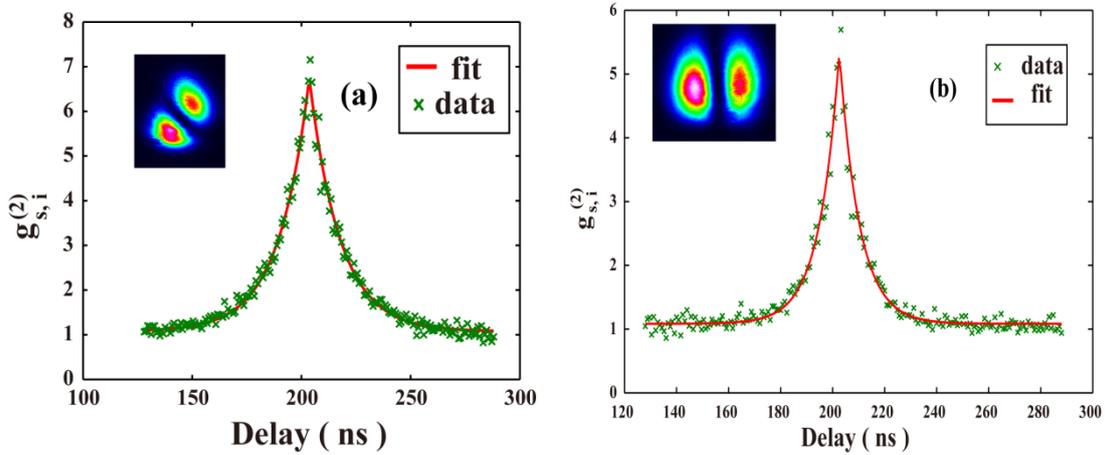

Figure 2. Normalized cross-intensity correlation between signal and idler photon. (a) the cavity mode is in diagonal direction; (b) the cavity mode is in horizontal direction.

Our experimental results clearly show that we could easily generate a photon pair non-classical correlated in different HG modes by locking the cavity in different HG modes. Besides, by using the type-I OPO operating far below threshold, we could also greatly reduce the bandwidth of the photon, which makes the effective coupling between the photon and atomic-based memory possible. The bandwidth of the photon pairs obtained experimentally are 11.4MHz and 20.8MHz for two different directions of the HG modes. In the present demonstration, the output spectral of the photon pair is in multi-longitude mode. Single longitude operation can be obtained by a filter. The filter methods introduced in ref. [30, 37] meet the requirements in our experiments. In the present experiments we aim to show the possibility to generate narrow bandwidth photon pairs in cavity's high order modes. This primary study will provide an effective way to generate HG mode narrow bandwidth photon pairs, which can be coupled to high dimensional quantum communications.

**Acknowledgements**

This work was supported by the National Fundamental Research Program of China (Grant No. 2011CBA00200), the National Natural Science Foundation of China (Grant Nos. 11174271,

**References**
[1] J. Wang, J.-Y. Yang, I. M. Fazal, N. Ahmed, Y. Yan, H. Huang, Y. Ren, Y. Yue, S. Dolinar, M. Tur, and A. E. Willner, "Terabit free-space data transmission employing orbital angular momentum multiplexing," Nature Photon. **6,** 488-496(2012).
[2] N. Bozinovic, Y. Yue, Y. Ren, M. Tur, P. Kristensen, H. Huang, A. E. Willner, and S. Ramachandran, Terabit-scale orbital angular momentum mode division multiplexing in fibers, Science 340, 1545-1548 (2013).
[3] G. Vallone, V. D'Ambrosio, A. Sponselli, S. Slussarenko, L. Marrucci, F. Sciarrino, and P. Villoresi1, Free-space quantum key distribution by rotation-invariant twisted photons, Phys. Rev. Lett. 113, 060503 (2014).
[4] S. Franke-Arnold, L. Allen, and M. Padgett, "Advances in optical angular momentum," Laser photon. Rev. 2, 299-313 (2008).
[5] V. D'Ambrosio, N. Spagnolo, L. Del Re, S. Sulssarenko, Y. Li, L. C. Kwek, L. Marrucci, S. P. Walborn, L. Aolita, and F. Sciarrinao, "Photonics polarization gears for ultra-sensitive angular measurements," Nature Commun. 4, 2432(2013).
[6]Z.-Y. Zhou, Y. Li, D.-S. Ding, Y.-K. Jiang, W. Zhang, S. Shi B.-S. Shi, and G.-C. Guo, "An optical fan for light beams for high-precision optical measurements and optical switching," Opt. Lett. 39, 5098-5101 (2014).
[7] M. P. J. Lavery, F. C. Speirits, S. M. Barnett, and M. J. Padgett, "Detection of a spinning object using light's orbital angular momentum," Science 341, 537-540 (2013).
[8] J. T. Barreiro, T.-C. Wei, and P. G. Kwiat, "Beating the channel capacity limit for linear photonic superdense coding," Nature Phys. 4, 282-286(2008).
[9] E. Nagali, L. Sansoni, F. Sciarrino, F. De Martini, L. Marrucci, B. Piccirillo, E. Karimi, and E. Santamato, "Optimal quantum cloning of orbital angular momentum photon qubits through Hong-Ou-Mandel coalescence," Nature Photon. 3, 720-723(2009).
[10] J. Leach, B. Jack, J. Romero, A. K. Jha, A. M. Yao, S. Franke-Arnold, D. G. Ireland, R. W. Boyd, S. M. Barnett, and M. J. Padgett, "Quantum correlations in optical angle-orbital angular momentum variables," Science 329, 662-665(2010).
[11] R. Fickler, R. Lapkiewicz, W. N. plick, M. Krenn, C. Schaeff, S. Ramelow, and A. Zeilinger, "Quantum entanglement of high angular momenta," Science 338, 640-643(2012).
[12]A. C. Dada, J. Leach, G. S. Buller, M. J. Padgett, and E. Andersson, "Experimental high-dimensional two-photon entanglement and violations of generalized Bell inequalities," Nat. Phys. 7, 677-680(2011).
[13]P. Kok, K. Nemoto, T. C. Ralph, J. P. Dowling, and G. J. Milburn, "Linear optical quantum computing with photonic qubits," Rev. Mod. Phys. 79, 135-174 (2007).
[14]J.-W. Pan, Z.-B. Chen, C.-Y. Lu, H.Weinfurter, A. Zeilinger, and M. Żukowski, "Multiphoton entanglement and interferometry," Rev. Mod. Phys. 84, 777-838 (2012).
[15]J. Barreiro, N. Langford, N. Peters, and P. Kwiat, "Generation of Hyperentangled Photon Pairs," Phys. Rev. Lett. 95, 260501 (2005).
[16]G. Vallone, R. Ceccarelli, F. De Martini, and P. Mataloni, "Hyperentanglement of two photons in three degrees of freedom," Phys. Rev. A 79, 030301(R) (2009).
[17]R. Ceccarelli, G. Vallone, F. De Martini, P. Mataloni, and A. Cabello, "Experimental

entanglement and nonlocality of a two-photon six-qubit cluster state," Phys. Rev. Lett. 103, 160401 (2009).

[18]J. T. Barreiro, T.-C. Wei, and P. G. Kwiat, "Beating the channel capacity limit for linear photonic superdense coding," Nature Phys. 4, 282-286(2008).

[19] D. S. Ding, Z.-Y. Zhou, B.-S. Shi, and G.-C. Guo, "Single-Photon-level quantum imaging memory based on cold atomic ensembles," Nat. Commun. 4, 2527 (2013);

[20]Dong-Sheng Ding, Wei Zhang, Zhi-Yuan Zhou, Shuai Shi, Jian-song Pan, Guo-Yong Xiang, Xi-Shi Wang, Yun-Kun Jiang, Bao-Sen Shi, and Guang-Can Guo, Toward high-dimensional-state quantum memory in a cold atomic ensemble," Phys. Rev. A. 90, 042301 (2014).

[21]A. Nicolas, L. Veissier, L. Giner, E. Giacobino, D. Maxein, and J. Laurat, "A quantum memory for orbital angular momentum photonic qubits," Nat. Photon. 8, 234-238 (2014).

[22]D.-S. Ding, W. Zhang, Z.-Y. Zhou, S. Shi, G.-Y. Xiang, X.-S. Wang, Y.-K. Jiang, B.-S. Shi, and G.-C. Guo, "Quantum storage of orbital angular momentum entanglement in an atomic ensemble," Phys. Rev. Lett. 114, 050502 (2015).

[23] V. Parigi, V. D'Ambrosioy, C. Arnoldy, L. Marrucci, F. Sciarrino, and J. Laurat, "Storage and retrieval of vector beams of light in a multiple-degree-of-freedom quantum memory," arXiv:1504.03096v1 [quant-ph].

[24]Z. Y. Ou, and Y. J. Lu, "Cavity enhanced spontaneous parametric down-conversion for the prolongation of correlation time between conjugate photons," Phys. Rev. Lett. 83, 2556-2559(1999).

[25]C. E. Kuklewicz, F. N. C. Wong, and J. H. Shapiro, "time-bin-modulated biphotons from cavity-enhanced down-conversion," Phys. Rev. Lett. 97, 223601(2006).

[26]F. Y. Wang, B. S. Shi, and G. C. Guo, "Observation of time correlation function of multi-mode two-photon pairs on a rubidium D2 line," Opt. Lett. 33, 2191-2193(2008); "Generation of narrow-band photon pairs for quantum momery" Opt. Commun. 283，2974, (2010).

[27] X. H. Bao, Y. Qian, J.Yang, H. Zhang, Z. B. Chen, T. Yang, and J. W. Pan, "Generation of narrow-band polarization-entangled photon pairs for atomic quantum memories," Phys. Rev. Lett. 101, 190501(2008).

[28] M. Scholz, L. Koch, and O. Benson, "Statistics of narrow-band single photons for quantum memories generated by ultra-bright cavity-enhanced parametric down-conversion," Phys. Rev. Lett. 102, 063603(2009).

[29] E. Pomarico, B. Sanguinetti, N. Gisin, R. Thew, H. Zbinden, G. Schreiber, A. Thomas, and W. Sohler, "Waveguide-based OPO source of entangled photon pairs," New J. Phys. 11, 113042(2009).

[30]F. Wolfgramm, Y. A. de I. Astiz, F. A. Beduini, A. Cere`, and M. W. Mitchell, "Atom-resonant heralded single photons by interaction-free measurement," Phys. Rev. Lett. 106, 053602 (2011).

[31] J. Fekete, D. Rieländer, M. Cristiani, and H. de Riedmatten, "Ultranarrow-Band Photon-Pair Source Compatible with Solid State Quantum Memories and Telecommunication Networks," Phys. Rev. Lett. 110, 220502 (2013).

[32] Z.-Y. Zhou, D.-S. Ding, Y. Li, F.-Y. Wang, and B.-S. Shi, "Cavity-enhanced bright photon pairs at telecom wavelengths with a triple-resonance configuration," J. Opt. Soc. Am. B 31, 128-134(2014).

[33]K. Liu, J. Guo, C. Cai, S. Guo, and J. Gao, "Experimental generation of continuous-variable hyperentanglement in an optical parametric oscillator," Phys. Rev. Lett. 110, 220502(2014).

[34] Y. Jeronimo-Moreno, S. Rodriguez-Bebavides, and A. B. U'Ren, "Theory of cavity-enhanced spontaneous parametric down conversion," Laser Physics 20, 1211-1233(2010).

[35] R. W. P. Drever, J. L. Hall, F. V. Kowalski, J. Hough, G. M. Ford, A. J. Munley, and H. Ward, "Laser phase and frequency stabilization using and optical resonator," Appl. Phys. B 31, 97-105(1983).

[36] A. Kuzmich, W. P. Bowen, A. D. Boozer, A. Boca, C. W. Chou, L. M. Duan, H. J. Kimble, "Generation of nonclassical photon pairs for scalable quantum communication with atomic ensembles," Nature 423, 731-734 (2003).

[37] M. A. Zentile, D. J. Whiting, J. Keaveney, C. S. Adams, and I. G. Hughes, "Atomic Faraday filter with equivalent noise bandwidth less than 1 GHz," Opt. Lett. 40, 200-2003(2015).